\documentclass[fleqn,usenatbib]{mnras}

\usepackage{newtxtext,newtxmath}

\usepackage[T1]{fontenc}

\DeclareRobustCommand{\VAN}[3]{#2}
\let\VANthebibliography\thebibliography
\def\thebibliography{\DeclareRobustCommand{\VAN}[3]{##3}\VANthebibliography}

\usepackage{graphicx}	
\usepackage{amsmath}	
\usepackage{enumitem}
\usepackage[utf8]{inputenc} 
\DeclareUnicodeCharacter{2212}{\ensuremath{-}} 
\usepackage{array}
\usepackage{booktabs}
\usepackage{multirow}
\usepackage{tabularx}
\usepackage{hyperref}

\title[Unsupervised Clustering of KAGRA O3GK Data]{Unsupervised Deep Learning Method for Clustering KAGRA O3GK Transient Noise Data}

\author[Y. Kim et al.]{
Yubin Kim,$^{1,4}$
Kihyun Jung,$^{2}$\thanks{E-mail: khjung@unist.ac.kr} and
Kyujin Kwak$^{2,3}$\thanks{E-mail: kkwak@unist.ac.kr}
\\
$^{1}$Department of Physics, Ewha Womans University, Seoul, 03760, Republic of Korea\\
$^{2}$Department of Physics, Ulsan National Institute of Science and Technology (UNIST), Ulsan, 44919, Republic of Korea\\
$^{3}$Graduate School of Artificial Intelligence, Ulsan National Institute of Science and Technology (UNIST), Ulsan, 44919, Republic of Korea\\
$^{4}$National Institute for Mathematical Sciences, Daejeon, 34047, Republic of Korea\\
}

\date{Accepted XXX. Received YYY; in original form ZZZ}

\pubyear{2026}

\begin{document}
\label{firstpage}
\pagerange{\pageref{firstpage}--\pageref{lastpage}}
\maketitle

\begin{abstract}
The advent of ground-based gravitational wave detectors has significantly improved the detection of faint gravitational wave signals.
However, these detectors are affected by various types of transient noise, known as glitches, which can mimic true gravitational wave signals and limit detector sensitivity. 
Classifying glitches according to their time-frequency characteristics not only facilitates a deeper understanding of their origins, but also supports their mitigation on data to maintain data quality.
While supervised machine learning methods are commonly employed for glitch classification,  they require the labelling of training datasets obtained through manual annotation, a process which is costly and not scalable for evolving detectors.  
In order to address this challenge, the present study investigates unsupervised deep learning methods for dimensionality reduction and clustering of glitch spectrogram images. 
In this study, three distinct approaches are applied and compared using data from the KAGRA detector's O3GK run.
The findings of this study demonstrate that deep learning-based feature extraction significantly enhances the clustering performance compared to traditional machine learning methods. 
This study presents an initial analysis of the KAGRA glitch dataset using unsupervised deep learning, highlighting the potential of this approach for efficient and scalable glitch classification in future observation runs.
\end{abstract}

\begin{keywords}
gravitational waves --- methods: data analysis --- software: machine learning 
\end{keywords}



\section{Introduction} \label{sec:intro}

The first observation of a gravitational wave signal in 2015 \citep{Abbott2016} ushered in the era of gravitational wave astronomy.
Gravitational wave detectors consist of highly sensitive laser interferometers and thousands of monitoring sensors, enabling the detection of extremely faint signals buried in noises. 
Improvements in detector sensitivity have increased the detection rate of gravitational wave signals.
However, ground-based detectors are challenged by transient noise, which limits their sensitivity.
Transient non-Gaussian noises, known as glitches, can lead to false alarms and affect the parameter estimation \citep{Abbott2018,Macas2022}. 
Therefore, identifying and removing them is crucial for ensuring data quality in the gravitational wave detection pipeline \citep{Covas2018}. 
Transient noise exhibits a variety of time-frequency patterns related to its source within the detector or in the environment.  
Classifying these glitches based on their time-frequency characteristics (i.e., the visual morphology observed in spectrograms) can provide insight into their origins and support efforts to improve detector sensitivity.

Machine learning provides an efficient tool for classifying a large number of glitches. 
While initial studies primarily utilised feature vectors computed from time series data \citep{Mukund2017, Powell2017}, the development of convolutional neural networks (CNNs) has shifted the focus towards image-based classification. 
In this approach, glitches are converted into two-dimensional spectrogram images using the Q-transform, enabling multi-class classification based on their distinct morphological features.
However, most existing machine learning approaches rely on supervised learning, which necessitates the extensive manual annotation of training data.  
The Gravity Spy project \citep{Zevin2017}, a pioneering effort in glitch classification, combined machine learning with citizen science to identify 22 classes of transient noise. The spectrogram images labelled by each class were used to train supervised machine learning models \citep{Bahaadini2018a}. 
However, as detector sensitivity improves, glitches become more numerous and diverse, making annotation increasingly costly. 
Furthermore, since glitches are dependent on the environment, new detector-specific training datasets are required for each new detector or upgrade, which limits scalability. 
Supervised algorithms also struggle to generalise beyond the fixed classes used in training, making it difficult to identify new glitches from detector upgrades \citep{Soni2021}.

In order to mitigate the reliance on annotated datasets in supervised learning, semi-supervised approaches have been proposed. 
Semi-supervised learning methods exploit unlabelled data together with a small set of labelled examples, alleviating annotation scarcity. 
For instance, \citet{Bahaadini2022} has reported on the training of a virtual adversarial model using a combination of labelled and unlabelled data. 
Another strategy is transfer learning, in which pre-trained CNNs are adapted for glitch classification.
\citet{Bahaadini2018b} introduced a feature extraction function based on a pre-trained VGG16 network to cluster glitches not belonging to any known Gravity Spy class, while \citet{George2018} showed that fine-tuned ImageNet models achieved high classification accuracy for glitch data. 
These studies further demonstrated that unsupervised clustering in the learned feature spaces can reveal novel glitch classes beyond existing categories.

In contrast to these approaches which still rely on some form of labelled data or prior knowledge, unsupervised learning does not require pre-assigned labels. 
Thus, it reduces annotation effort and enables the identification of novel glitch types.
Recent studies have demonstrated the potential of unsupervised learning for LIGO glitch analysis. 
For instance, \citet{Sakai2022} proposed a framework combining a variational autoencoder (VAE) with invariant information clustering for Gravity Spy O1 data, while \citet{Li2024} presented an autoencoder-based architecture integrating CNNs and vision transformers to reduce dimensionality and cluster Gravity Spy O3 data. 
However, most existing glitch classification studies have focused on the LIGO, while the glitch properties of other detectors, such as Virgo and KAGRA, have received comparatively little attention. 
As KAGRA has joined recent observing runs, the accurate characterisation of the detectors has become increasingly significant. 
Nevertheless, only a few studies have used deep learning techniques to analyse KAGRA glitches. 
\citet{Shoichi2025} performed unsupervised clustering and proposed the optimal number of clusters in the KAGRA glitch data during the O3GK run period by combining a VAE and spectral clustering. 

This study advances the unsupervised analysis of glitch characteristics by focusing on a subset of total glitch events whose correlations with auxiliary channels have been identified.
Unlike previous studies, a systematic comparison of multiple unsupervised learning approaches is conducted to evaluate their validity and performance.  
In particular, deep learning models are incorporated into the dimensionality reduction stage to capture the intrinsic features of spectrogram images more effectively, thereby improving the performance of the clustering process.
The objective of this comparative analysis is to identify unsupervised methods that achieve robust and optimal clustering performance for the KAGRA glitch data. 
As future observatories integrate into the global network, unsupervised clustering methods will play a crucial role in identifying novel glitch classes without prior labelled data.
Therefore, this work aims to address the existing gap in glitch analysis by presenting the unsupervised analysis of KAGRA O3GK data. 
It also demonstrates the potential of deep learning-based clustering methods as scalable tools for detector characterisation.

\section{Dataset} \label{sec:dataset}

This section details the acquisition, pre-processing, and structural framework of the KAGRA O3GK dataset. It also describes the composition and distribution of the glitch samples used as the input for our proposed model.

\subsection{KAGRA O3GK Dataset}\label{sec:dataset1}

\begin{figure*}
    \centering
    \includegraphics[width=0.85\linewidth]{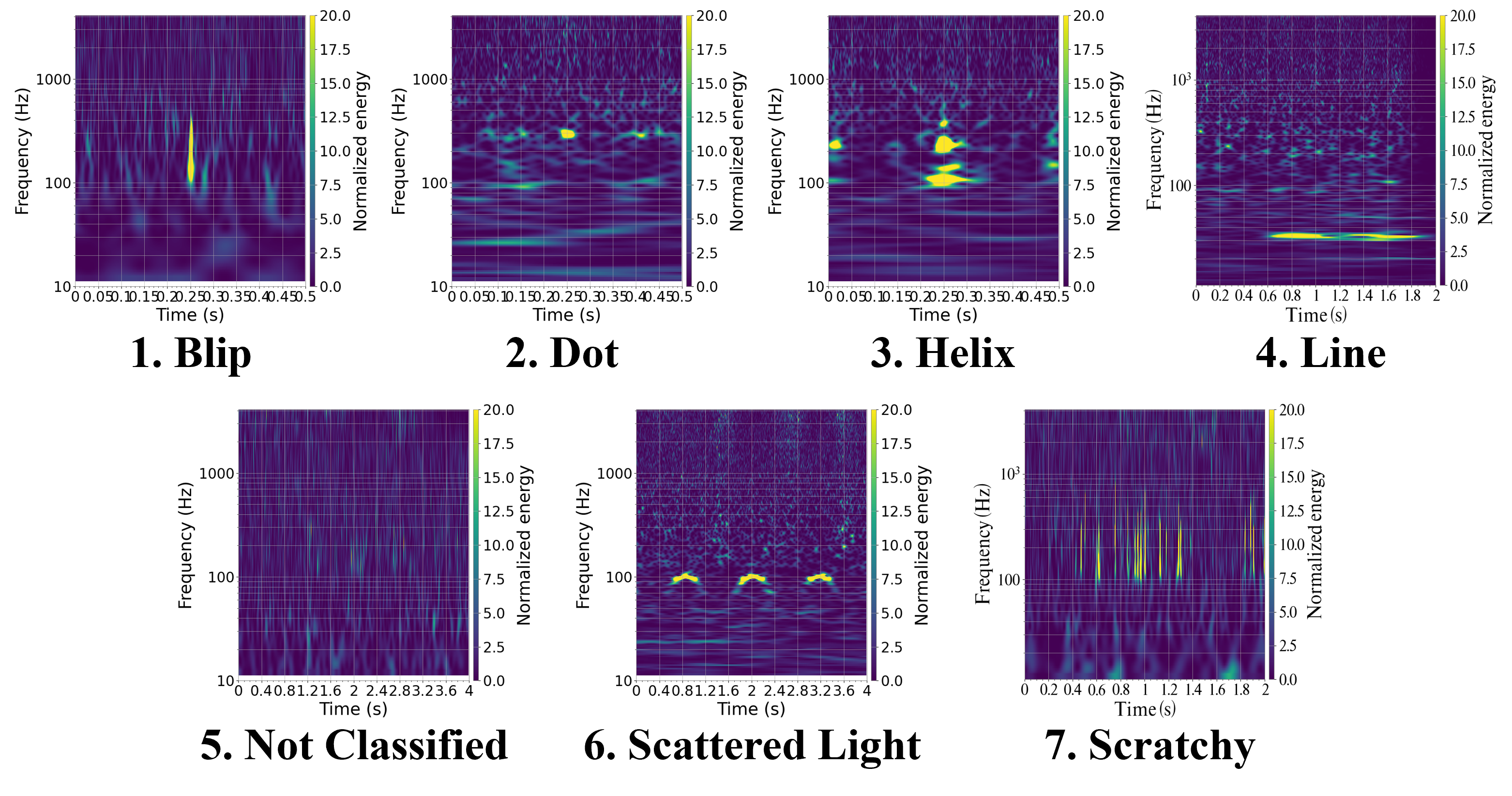}
    \caption{Example spectrogram images of glitches used in this study, shown for each glitch class. All images correspond to glitch events identified by Hveto and annotated by human experts. More detailed descriptions of each glitch class can be found in \citet{Akutsu2025}.}
    \label{fig1}
\end{figure*}

KAGRA \citep{Akutsu2019}, the gravitational wave detector located in the Kamioka mine in Japan, is uniquely operated underground with cryogenic mirrors to reduce thermal noise. 
Although KAGRA did not fully participate in the third observing run (O3) of LIGO and Virgo, it conducted its first joint observing run, named O3GK, in collaboration with the German--British detector GEO600 \citep{KAGRA2022}. 
This run lasted for approximately two weeks, from April 7, 2020, 08:00 UTC (GPS time: 1270281618) to April 21, 2020, 00:00 UTC (GPS time: 1271462418). 

In this study, we use glitches in the KAGRA O3GK data that were identified in \citet{Akutsu2025}. Here we briefly summarise the process used in \citet{Akutsu2025}. First, we generate triggers for both main channel (strain data) and auxiliary channels (data from noise-monitoring systems) by using Omicron \citep{ROBINET2020}. Omicron detects excess power in the time-frequency representation using Q-transform. Each detection is assigned a signal-to-noise ratio (SNR) and recorded as a trigger. The hierarchical veto (Hveto; \citet{Smith2011}) is subsequently applied to statistically evaluate the temporal coincidence between the triggers in the main channel and those in the auxiliary channels. While the primary function of Hveto is to prevent low-quality data from being utilised in the analysis and to enhance the sensitivity of gravitational wave searches, the process in \citet{Akutsu2025} utilised its output to specifically isolate glitches with high probability of instrumental origin based upon statistical time correlation. This approach differs from the study by \citet{Shoichi2025}, which analysed the same O3GK main channel data but did not account for auxiliary channel correlations via Hveto. Consequently, the filtered subset of high-confidence instrumental glitches used in this work (2,531 events) is lower than the total number of glitches in \citet{Shoichi2025} (45,345 events). 


This study utilised a dataset of two-dimensional spectrogram images of glitches observed in the main channel of KAGRA during the O3GK run, from 7 to 21 April 2020. 
Although we used auxiliary channel data when applying Hveto, a dataset of two-dimensional spectrogram images of glitches for this study are produced only from the main channel (strain data). 
We note that the main channel was K1:DAC-STRAIN\_C20, while auxiliary channels belonged to eight subsystems: AOS, CAL, IMC, LAS, LSC, PEM, PSL, and VIS.
By applying an SNR threshold of 8 to the main channel, Hveto identified 2,531 vetoed events across 28 auxiliary channels.
Spectrogram images of these vetoed events, corresponding to glitches in the main channel, were generated using the Q-transform module implemented in GWpy \citep{MACLEOD2021}. 

In \citet{Akutsu2025}, the resulting 2,531 glitch events were manually grouped into six classes based on the morphological characteristics of excess power area in the spectrograms: blip, dot, helix, line, scattered light, and scratchy.
Blip, helix, scratchy, and scattered light glitches are similar in morphology to existing classes in LIGO datasets, while dot and line glitches were newly defined for KAGRA, as they are not included in the Gravity Spy taxonomy \citep{Zevin2017}. 
During the labelling process, events that did not fit into these six categories were assigned to a ``Not Classified" category.
These particular events show no apparent transient noise structure in the main channel spectrogram, even though Hveto identified a statistical correlation with the auxiliary channels. 
These glitches might be visibly shrunk to the point that they blend into the background because of the logarithmic frequency scale. 
In previous studies of glitch classification based on Gravity Spy taxonomy, such spectrograms have typically been categorised as ``No Glitch" \citep{Bahaadini2018a,Glanzer2023}.
Rather than excluding these ``Not Classified" events from the dataset, we retained them as a distinct category to evaluate the capability of our unsupervised model in distinguishing this lack of clear morphological patterns from other glitch classes.
As a result, the 2,531 vetoed event spectrograms were categorised into the following seven classes: blip, dot, helix, line, not classified, scattered light, and scratchy. 
Representative examples of each glitch morphology, along with their assigned labels, are shown in \autoref{fig1}.

\subsection{Input Data Pre-processing}\label{sec:dataset2}

The input data utilised in this study consist of the glitch spectrogram images identified in the KAGRA main channel during the O3GK run, as outlined in the preceding section.
These spectrograms were subjected to a series of pre-processing steps prior to being entered as inputs into the unsupervised clustering models.
For each confirmed glitch event from the Hveto result, spectrogram images corresponding to four different time-scales (0.5, 1.0, 2.0, and 4.0 seconds) were generated, centred on the event time.
During this process, the events located at the boundaries of data quality flags were excluded from the input dataset. 
Such segments are subject to edge effects or processing artefacts, rendering them unsuitable for scientific analysis.
Out of the initial 2,531 glitch events, 316 such events were visually identified on the spectrograms and removed; the four time-scale spectrograms for the remaining 2,215 selected events were then subjected to additional pre-processing steps.

\begin{figure*}
    \centering
    \includegraphics[width=0.85\linewidth]{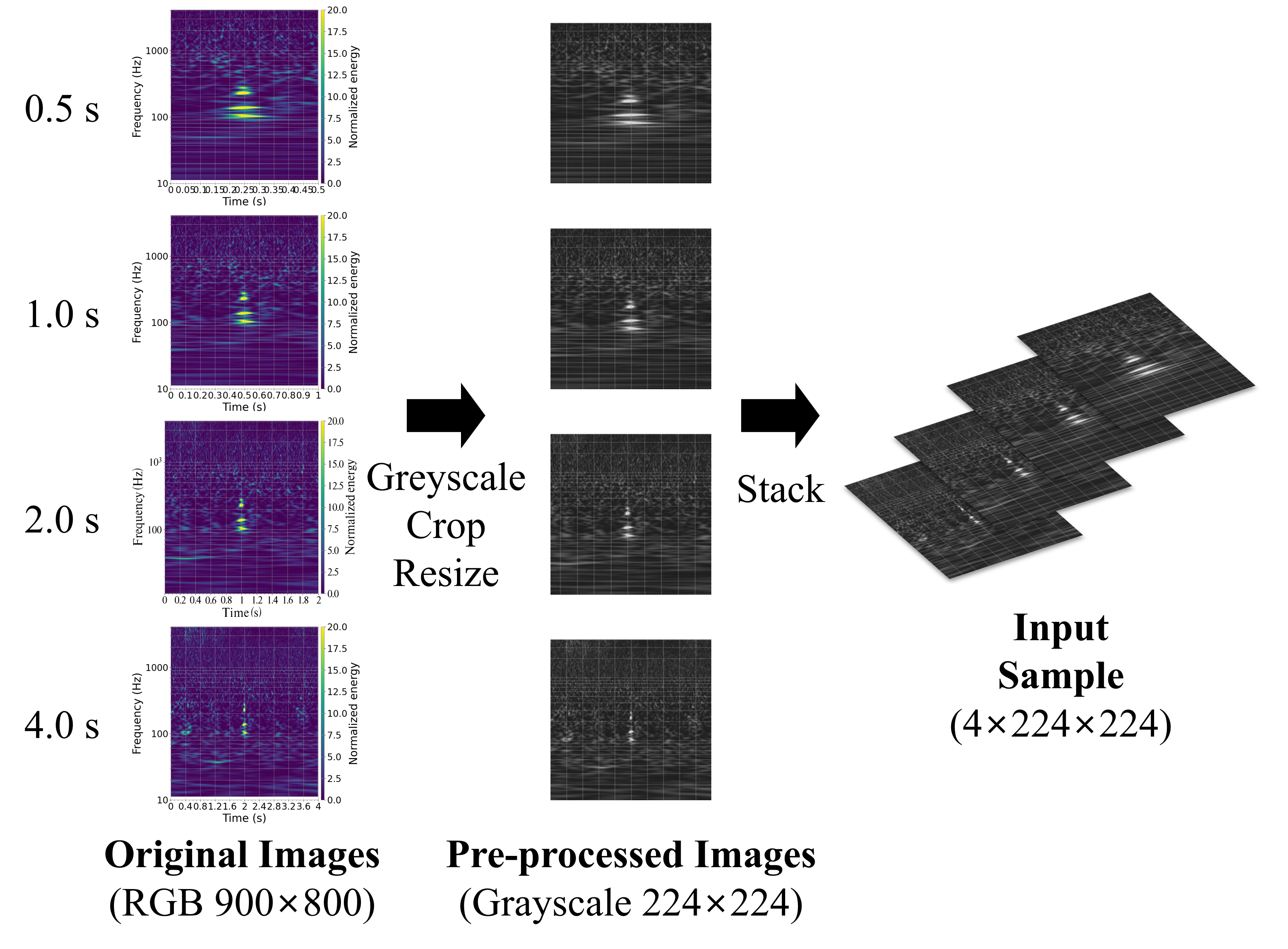}
    \caption{Overview of the input data pre-processing. For each glitch event, spectrogram images in four different time-scales were cropped to retain only the spectrogram region and resized into square greyscale images. These four images were then stacked to form a single four-channel input sample.}
    \label{fig2}
\end{figure*}

The original spectrogram images were initially represented as RGB images with a resolution of $900 \times 800$ pixels. 
Each image was first converted into a single-channel greyscale representation. 
Subsequently, the images were cropped to remove the titles and axis labels, leaving only the spectrogram content. 
The resulting cropped region comprised $540 \times 624$ pixels and was then resized to a standardised square format of $224 \times 224$ pixels. 
Through this pre-processing steps, we can reduce the computational cost by decreasing the dimensionality of the input data while preserving the essential morphological features required for unsupervised feature extraction and clustering.
Converting to greyscale and resizing may slightly alter the morphology or energy scaling of the original spectrograms, but its effect on our current study is not significant because this transformation is applied consistently across all samples. 
We note that the primary objective of our study is not to precisely reconstruct physical parameters from these images but rather to automate the grouping of morphologically similar glitches.
For each glitch event, spectrogram images corresponding to four different time durations were stacked together to form the final input representation. 
This multi-scale tensor ($4 \times 224 \times 224$) enables our unsupervised machine learning models to capture features across different time-frequency resolutions simultaneously.
While the physical time step per pixel varies across channels, all spectrograms are strictly centre-aligned with respect to the exact glitch event time.
An overview of the pre-processing workflow is shown in \autoref{fig2}. 

After pre-processing, a total of 2,215 four-channel stacked image samples were obtained for the experiments.
The dataset was split into three sets: a training set comprising 70\% of the total data, a validation set consisting of 10\%, and a test set containing the remaining 20\%. 
In consideration of the highly imbalanced distribution of glitch samples across the defined morphological classes, we employed a stratified split for data partitioning. 
This method ensures that the original class proportions are strictly preserved within both the training and validation subsets. 
By preventing minority glitch classes from being under-represented or omitted during the training phase, the stratified partition enables the model to effectively capture the intrinsic structural and morphological features of all glitch types.
The pre-assigned labels from the manual classification were solely used as ground truth for post-hoc performance evaluation and were not involved in model training process.
The distribution of samples across glitch classes for each subset is summarised in \autoref{table1}.

\begin{table}
\centering
\begin{tabular}{cccc}
\hline
\textbf{Class} & \textbf{Train} & \textbf{Val} & \textbf{Test} \\ \hline
Blip           & 816            & 117          & 233           \\
Dot            & 33             & 5           & 9             \\
Helix          & 200            & 28           & 57            \\
Line           & 304            & 44           & 87            \\
NotClassified  & 62             & 9           & 18            \\
ScatteredLight & 37             & 5            & 11            \\
Scratchy       & 98            & 14            & 28            \\ \hline
Total          & 1550           & 222          & 443           \\ \hline
\end{tabular}
\caption{Distribution of input samples across the seven glitch classes in the training, validation, and test sets.}
\label{table1}
\end{table}

\section{Methods} \label{sec:method}

In this section, we present a description of the approaches used in our study to perform unsupervised clustering on KAGRA O3GK glitch data, the experimental design, and the evaluation metrics used to compare experimental results obtained with different methods.

\subsection{Unsupervised Machine Learning Method}\label{sec:method1}

The direct application of clustering algorithms to the raw pixels of spectrogram images is impractical.
In high-dimensional feature spaces, data points become extremely sparse, and the distance metrics used for clustering lose their discriminative power; this ultimately degrades model performance and increases computational costs.
Therefore, dimensionality reduction is a crucial step in unsupervised clustering, as it maps high-dimensional spectrogram data into a compact latent space while preserving the most informative features.
This process suppresses noise and emphasizes meaningful structures prior to clustering, enabling more accurate and computationally efficient grouping of glitch events.
Principal component analysis (PCA) \citep{Pearson1901} is a well-established method for linear dimensionality reduction. PCA eliminates redundant dimensions by projecting data onto a set of orthogonal components that preserve dominant variance. For example, an image of size 900 × 800 contains 720,000 dimensions, which can be reduced to a low-dimensional representation (e.g., 20--50 components) using PCA. In this study, PCA-reduced feature vectors combined with \textit{k}-means clustering \citep{Macqueen1967} are adopted as a baseline method.

In order to capture more complex and non-linear structures in glitch spectrograms, we employ deep learning-based dimensionality reduction methods involving a convolutional autoencoder (ConvAE) \citep{Masci2011}.
The resulting latent representations are subsequently clustered using \textit{k}-means clustering or deep embedded clustering (DEC) \citep{Xie2016}. 
Deep learning-based approaches enhance our understanding of the intrinsic structure of the high-dimensional data, ultimately leading to improved glitch classification. The following subsections describe the detailed algorithms of the machine learning methods employed in the experiment.

\subsubsection{Convolutional Autoencoder}

\begin{figure*}
    \centering
    \includegraphics[width=0.9\linewidth]{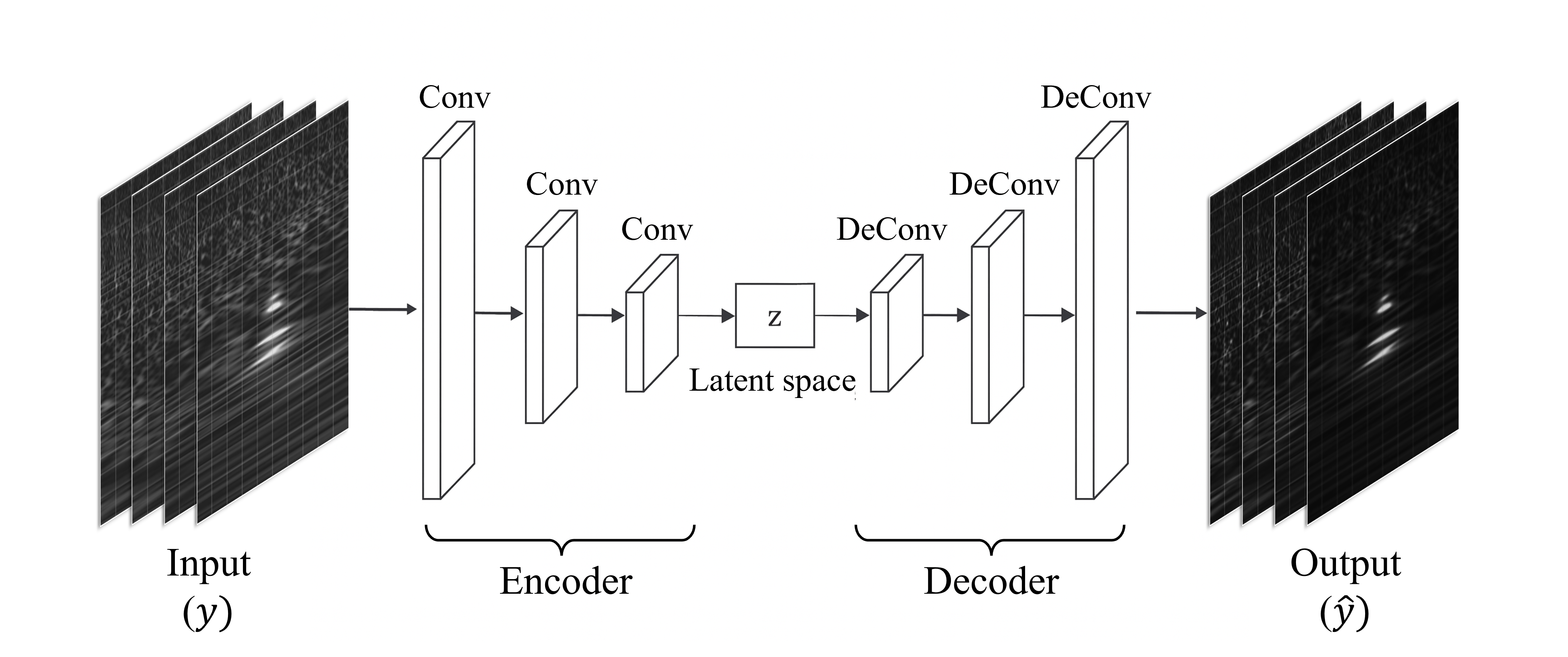}
    \caption{Schematic illustration of the ConvAE architecture. }
    \label{fig3}
\end{figure*}

An autoencoder \citep{Hinton2006} is an unsupervised neural network designed to learn a compressed, low-dimensional representation of input data. 
It consists of two main components: an encoder which maps high-dimensional input to a lower-dimensional latent space; and a decoder which reconstructs the original data from this latent representation.
A ConvAE is specifically designed autoencoder for image data with spatial structure \citep{Masci2011}. 
Instead of using fully connected layers, the ConvAE uses CNNs in both the encoder and decoder. 
This architecture allows the model to effectively capture and preserve the spatial hierarchies and local features of the input images, making it highly efficient for image-based tasks and non-linear mapping of the data. 

\autoref{fig3} illustrates the architecture of a ConvAE. The encoder maps the input $y \in \mathbb{R}^{H \times W \times C}$ into a low-dimensional latent representation $z$, while the decoder reconstructs the input from this latent code:
\[
z = f_\phi(y), \qquad \hat{y} = g_\psi(z).
\]
Here, $f_\phi$ and $g_\psi$ denote the encoder and decoder functions, respectively, with learnable parameters $\phi$ and $\psi$.
The encoder is composed of successive convolutional and down-sampling operations that progressively reduce the spatial resolution while extracting higher-level features. 
The latent representation $z$ serves as a compressed embedding that captures the most salient information of the input. 
The decoder mirrors this structure, employing up-sampling and convolutional operations to reconstruct the input from $z$.

Training of the ConvAE is performed by minimizing the difference between the input $y$ and its reconstruction $\hat{y}$. 
In this work, the mean squared error (MSE) is employed as the reconstruction loss: 
\begin{equation}
\mathcal{L}_{\mathrm{MSE}} = \frac{1}{N} \sum_{i=1}^{N} \| y_i - \hat{y}_i \|^2,
\label{eq:mse_loss}
\end{equation}
where $N$ is the number of training samples. 
This process encourages the autoencoder to learn latent features that preserve the essential structure of the data while discarding noise and redundancy. 
The detailed training process and hyperparameters used in this study will be descried in \autoref{sec:result1}.

\subsubsection{\textit{k}-means Clustering}

\textit{k}-means clustering is one of the most widely used clustering algorithms that partitions unlabelled data into $K$ distinct clusters. The algorithm begins by initialising $K$ centroids, after which each data point is assigned to the nearest centroid. The cluster centroids are then updated by computing the mean of all the points assigned to each cluster. These assignment and update steps are repeated iteratively until the centroids no longer change. 
Due to its computational simplicity and efficiency, \textit{k}-means clustering is widely applied in practice. 
However, the method is sensitive to centroid initialisation and requires the number of clusters $K$ to be specified in advance, which can limit its flexibility.

Let $\mathcal{X}=\{x_i\}_{i=1}^{N}\subset\mathbb{R}^d$ denote the dataset, where $N$ is the number of data points and $d$ is the feature dimension. Let $K$ be the number of clusters, $\mu_j \in \mathbb{R}^d$ the centroid of cluster $j$, and $c_i \in \{1,\dots,K\}$ the cluster index assigned to $x_i$.
The objective of \textit{k}-means is to partition the observations into $K$ sets so as to minimize the within-cluster sum of squares, which is the total sum of the squared Euclidean distances between each data point and the centroid of its assigned cluster given as:
\begin{align}
J\big(\{\mu_j\},\{c_i\}\big)
  &= \sum_{i=1}^{N} \left\| x_i - \mu_{c_i} \right\|^2.
  \label{eq3}
\end{align}
Given the current centroids, each data point is assigned to the nearest centroid:
\begin{align}
c_i
  &= \arg\min_{j\in \{1,\dots,K\}} \left\| x_i - \mu_j \right\|^2.
  \label{eq4}
\end{align}
Given the current assignments, each centroid is updated by the mean of its cluster:
\begin{align}
C_j &= \{\, x_i \in \mathcal{X} \mid c_i = j \,\}, \\
\mu_j
  &= \frac{1}{|C_j|} \sum_{x_i \in C_j} x_i, 
  \label{eq5}
\end{align}
where $|C_j|$ denotes the cardinality of the set $C_j$, representing the number of data points assigned to cluster $j$. 
The assignment step \eqref{eq4} and the update step \eqref{eq5} are iterated
until convergence, i.e., no change in $\{c_i\}$.


\subsubsection{Deep Embedded Clustering}
DEC is an unsupervised clustering algorithm that integrates the feature extraction with clustering, enabling the joint optimisation of latent representations and cluster assignments. 
Initially, a deep autoencoder with multiple hidden layers is trained to map input data into a low-dimensional latent space. 
The initial cluster centroids are obtained by applying \textit{k}-means clustering to the latent features, which are low-dimensional vectors embedded in latent space. 
Following initialisation, the encoder is subsequently fine-tuned to sharpen cluster assignments by enhancing intra-cluster compactness and inter-cluster separation. 

The detailed optimisation process of DEC is described as follows.
First, the model computes the soft assignments, which represent the probability that each sample belongs to a given cluster.
The similarity between an embedded point $z_i$ and a cluster centroid $\mu_j$ is measured using Student’s t-distribution. 
The resulting probability (soft assignment) of assigning sample $i$ to cluster $j$ is denoted by $q_{ij}$, which is formally defined as:
\begin{equation}
q_{ij} = 
\frac{ \left( 1 + \| z_i - \mu_j \|^2 / \alpha \right)^{-\frac{\alpha+1}{2}} }
     { \sum_{j'} \left( 1 + \| z_i - \mu_{j'} \|^2 / \alpha \right)^{-\frac{\alpha+1}{2}} },
\label{eq:qij}
\end{equation}
where $z_i = f_\theta(x_i) \in Z$ corresponds to the embedded representation of $x_i \in X$, and 
$\alpha$ is the degree of freedom of the Student's $t$-distribution which determines the kurtosis of the distribution. 
Second, a target distribution $p_{ij}$ is then constructed by squaring and normalising the soft assignments $q_{ij}$,
\begin{equation}
p_{ij} = 
\frac{ q_{ij}^2 / f_j }
     { \sum_{j'} q_{ij'}^2 / f_{j'} },
\label{eq:pij}
\end{equation}
where $f_j = \sum_i q_{ij}$ represents the soft cluster frequencies, which act as normalisation factors that scale the loss contribution of each cluster based on its total current size.
Finally, the model is trained by minimizing the Kullback--Leibler (KL) divergence between the target distribution $P$ and the soft assignment distribution $Q$,
 \begin{align}
     KL(P||Q) = \underset{i}{\sum}\underset{j}{\sum} p_{ij}\log \frac{p_{ij}}{q_{ij}}.
\end{align}
Through this process, the parameters of the deep neural network and the set of \textit{k} cluster centroids in the latent space $z$ are updated jointly. Owing to its ability to capture non-linear structures, DEC is well-suited for complex, high-dimensional data and has demonstrated superior performance compared to conventional clustering methods across multiple benchmark datasets \citep{Xie2016, Guo2021}.

\subsection{Experiment Design}\label{sec:method2}

\begin{figure}
    \centering
    \includegraphics[width=0.9\linewidth]{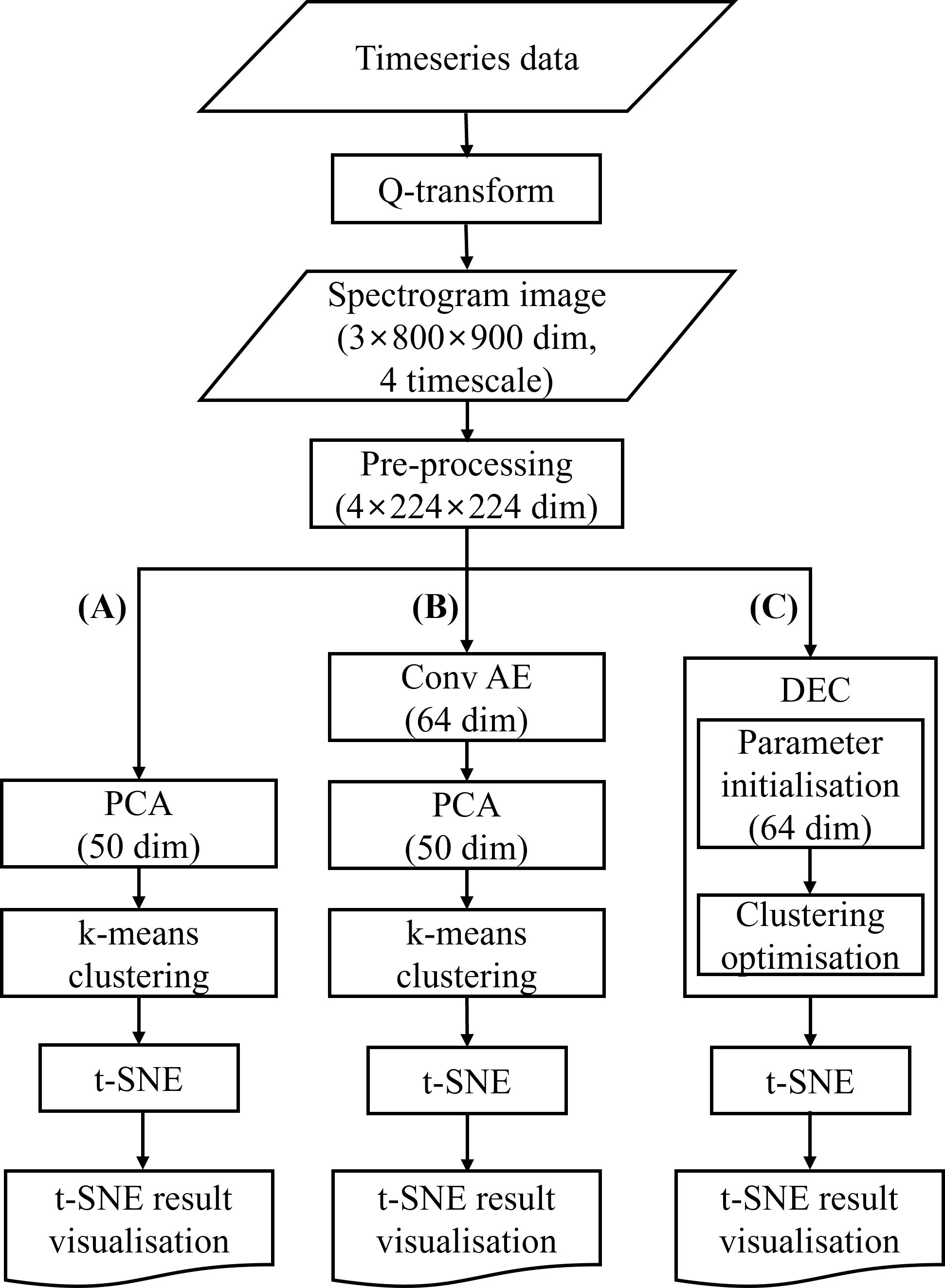}
    \caption{Workflow of data pre-processing and unsupervised clustering experiments applied to KAGRA O3GK glitch dataset. After spectrogram pre-processing, three distinct methods are applied: (A) PCA followed by \textit{k}-means clustering; (B) ConvAE-based feature extraction combined with PCA and \textit{k}-means clustering; and (C) DEC algorithm, which jointly optimises feature embedding and cluster assignments. The resulting clusters are visualised using t-SNE for comparison.}
    \label{fig4}
\end{figure}

In this study, three unsupervised clustering approaches are presented and compared.
\autoref{fig4} shows the schematic workflow of our study. The first four steps relevant to data preparation and pre-processing were explained in \autoref{sec:dataset}. The subsequent steps are roughly divided into two stages (dimensionality reduction and clustering) and three different unsupervised clustering approaches are adopted and compared. 
In approaches (A) and (B), PCA and ConvAE are employed to perform dimensionality reduction, followed by a separate clustering stage using the \textit{k}-means algorithm. 
In contrast, approach (C) employs DEC, where dimensionality reduction and clustering are jointly optimised within a unified framework. 
In order to facilitate visual interpretation and comparison of the clustering outcomes across the three approaches, t-distributed stochastic neighbour embedding (t-SNE) \citep{Laurens2008} is employed. 
This technique projects high-dimensional data into two dimensions while preserving local structural relationships, enabling the intuitive visualisation of the clustering results. 
The subsequent paragraphs describe each approach in detail.

\begin{enumerate}[label=(\Alph*)]
    \item PCA + \textit{k}-means: Using PCA for dimensionality reduction followed by \textit{k}-means clustering is a common approach in unsupervised clustering. In this study, this common approach is adopted as the baseline method. Initially, each four-channel input image was standardised and reduced to a 50-dimensional vector via PCA. Subsequently, the $k$-means algorithm was applied to the reduced 50-dimensional feature vectors to partition the dataset into a predefined number of clusters ($K = 7$), which was chosen from the seven glitch classes identified in our dataset (\autoref{table1}).
    
    \item ConvAE + PCA + \textit{k}-means: While PCA provides a simple linear baseline, it may not fully capture the intrinsic structure of high-dimensional spectrogram data. To address this limitation, the ConvAE maps input images into 64-dimensional latent vectors prior to the application of PCA. Consistent with the baseline approach, these latent vectors from encoder were standardised before applying PCA. The latent vectors compressed by the encoder were further reduced to 50 dimensions using PCA and subsequently clustered into seven clusters using the \textit{k}-means algorithm, as in the baseline method. The architecture and details of the ConvAE model employed in this study are summarised in \hyperref[sec:appendix]{Appendix}.
    
    \item DEC: DEC algorithm involves the parameter initialisation and joint optimisation of the latent representation and cluster assignments. The same ConvAE architecture as in (B) is used to map the input spectrogram images into a 64-dimensional latent space, providing an initial low-dimensional representation of the data. Initial cluster centroids are then obtained in this latent space using \textit{k}-means clustering. During subsequent training, DEC updates the encoder parameters by refining the latent representations to improve the resulting cluster assignments.

    \end{enumerate}

\subsection{Evaluation Metric}\label{sec:method3}

\subsubsection{Clustering Quality Evaluation}

The separation of the clusters in the experimental results was quantitatively assessed by computing the silhouette value. The silhouette value is a widely used internal clustering validation metric that simultaneously measures cluster cohesion (intra-cluster similarity) and separation (inter-cluster distance) based on pairwise distances between data points. Unlike external metrics, it evaluates clustering quality independently of ground-truth labels.

For a given data point $i$, let $d(i, j)$ denote the distance between $i$ and the other point $j$. Then the average distance between data points $i$ and $j$ belonging to the same cluster $C_I$ is given by
\begin{align}
    a(i)=\frac{1}{|C_I|-1}\underset{i,j \in C_I, j \ne i}{\sum}d(i,j),
\end{align}
where $|C_I|$ is the number of data belonging to cluster $C_I$. 
The average distance between data point $i$ belonging to cluster $C_I$ and other data points $j$ belonging to the nearest other cluster $C_J$ is given by
\begin{align}
    b(i)=\underset{J\ne I}{\min} \frac{1}{|C_J|} \underset{i \in C_I, j \in C_J}{\sum}d(i,j).
\label{eq:bi}
\end{align}
Using $a(i)$ and $b(i)$, we can calculate similarity between two nearest clusters.
The silhouette value $s(i)$ for data point $i$ is defined as
\begin{align}
    s(i) = 
    \frac{b(i)-a(i)}{\max \{a(i), b(i)\}}.  
\end{align}
The silhouette value ranges from −1 to 1. Positive values indicate appropriate cluster assignment, while values close to zero suggest that a data point lies near the boundary between clusters. 
Negative values imply that the sample may have been incorrectly assigned to a cluster. 
The mean silhouette value across all data points in the dataset, referred to as the silhouette score, is used as an overall measure of clustering quality. 
Clustering quality such that the silhouette score being close to 1 indicates distinctive separation among different clusters.

\subsubsection{Clustering Performance Evaluation}

To evaluate the performance of the clustering algorithms, two widely utilised external validation metrics were employed: normalised mutual information (NMI) and adjusted rand index (ARI). Both metrics are external indices that require ground-truth labels and quantify the degree to which the predicted clusters align with the true class distribution \citep{Nguyen2010}.
NMI and ARI are considered robust metrics for unsupervised learning, as they remain applicable and reliable even when the number of predicted clusters does not align with the number of ground-truth classes.

In this study, we calculate the NMI value with the default implementation in the \texttt{scikit-learn} library\footnote{\scriptsize\url{https://scikit-learn.org/stable/modules/generated/sklearn.metrics.normalized_mutual_info_score.html}}given as
\begin{equation}
\mathrm{NMI}(Z; \hat{Z}) = \frac{2 I(Z; \hat{Z})}{H(Z) + H(\hat{Z})},
\label{eq:nmi}
\end{equation}
where the arithmetic mean is used for the normalization factor. In the above equation, $Z$ and $\hat{Z}$ are the ground-truth and predicted cluster, respectively, and $H(U)$ is the entropy of a given cluster $U$ which is calculated based upon fundamental concepts in the information theory \citep{Wiley2005}.
In the context of unsupervised learning, high (low) value of H(U) implies that U contains a large (small) number of different classes into which data are classified. $I(U,V)$ is mutual information (MI) between two clusters U and V, given as $I(U,V) = H(V) - H(V|U)$ where $H(V|U)$ is conditional entropy which is the weighted average of the entropies of V for each specific value that U can take. MI measures the amount of information that U and V share and indicates the reduction in uncertainty of U due to the knowledge of V. We note that NMI value ranges from 0 (no mutual information) to 1 (perfect agreement).



ARI evaluates clustering similarity by considering all pairs of samples and counting pairs that are correctly assigned to the same or different clusters in both the predicted and ground-truth labels. ARI is a chance-adjusted version of the Rand Index (RI) which measures the proportion of correctly grouped pairs. The RI is defined as 
\begin{equation}
\mathrm{RI} = \frac{C_1 + C_2}{\binom{n}{2}},
\label{eq:ri}
\end{equation}
where $n$ is the total number of samples, $C_1$ is the number of pairs correctly placed in the same cluster in both ground truth and prediction, and $C_2$ is the number of pairs correctly separated into different clusters by both. Unlike RI, ARI accounts for random agreement, and its value may take negative values when the clustering result is worse than random assignment. ARI is defined as
\begin{equation}
\mathrm{ARI} = \frac{\mathrm{RI} - E[\mathrm{RI}]}{\max(\mathrm{RI}) - E[\mathrm{RI}]},
\label{eq:ari}
\end{equation}
where $E[\mathrm{RI}]$ and $\max(\mathrm{RI})$ are the expected value and maximum value of RI, respectively. 
The ARI ranges from $-1$ to $1$: $1$ indicates perfect agreement between the clustering assignments while $0$ corresponds to random distribution. Negative values indicate agreement worse than random.

\section{Results}


This section presents the clustering outcomes obtained through the three independent unsupervised approaches detailed in \autoref{sec:method2}. 
We first outline the training progress and hyperparameters of the deep learning models, followed by a comparative analysis of their respective feature extraction and clustering performances.

\subsection{Training Process of Deep Learning Model}\label{sec:result1}

\begin{table}
\centering
\begin{tabular}{ccc}
\hline
                     & Hyperparameter            & Value \\ \hline
Training   of ConvAE & Learning rate             & 0.001 \\
                     & Mini-batch   size              & 32    \\
                     & Latent   vector dimension & 64    \\
                     & Epoch   number            & 50    \\ \hline
Training   of DEC    & Cluster   number          & 7     \\
                     & Convergence   tolerance   & 0.001 \\
                     & Alpha                     & 1.0   \\ \hline
\end{tabular}
\caption{Hyperparameters for training ConvAE and DEC.}
\label{table3}
\end{table}

\begin{figure}
    \centering
    \includegraphics[width=0.85\linewidth]{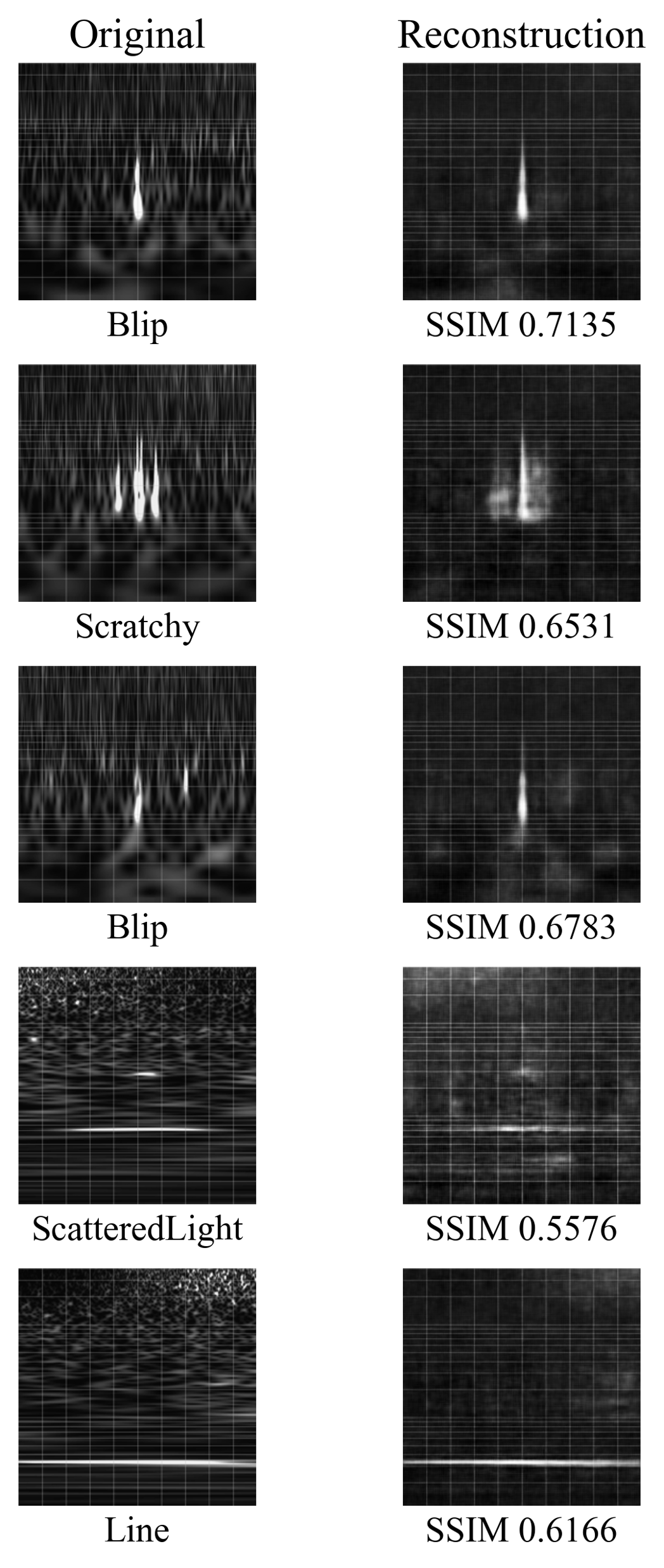}
    \caption{Example input and output of the ConvAE model. The left column shows input spectrogram images at the 0.5-second time-scale for representative glitch classes, while the right column shows the corresponding reconstructions produced by the ConvAE decoder. The similarity index is reported for each reconstructed image to quantify similarity to the input.}
    \label{fig5}
\end{figure}

\autoref{table3} summarises the parameters utilised for training the ConvAE and DEC models.
A detailed description of the ConvAE architecture is provided in \hyperref[sec:appendix]{Appendix}.
The dataset was partitioned into mini-batches of size 32 to update the weights of the neural network, with the network parameters updated upon completion of each individual batch.
We employed the Adam optimiser \citep{Diederik2017} with an initial learning rate of $0.001$, an algorithm that adaptively adjusts the learning rate for each parameter. 
The ConvAE was trained over 50 epochs, a duration found to be sufficient for the reconstruction loss to stabilise without overfitting. 
Additionally, batch normalisation was applied after each convolutional layer to rescale the feature maps and stabilise the learning process.
The effectiveness of this training is qualitatively demonstrated in \autoref{fig5}, which illustrates a series of input image samples and their corresponding reconstructions produced by the decoder.
Higher similarity between the input images and the reconstructed images indicates that the latent vectors effectively preserve the essential features of the original images during compression. 
To quantitatively assess the similarity between the original and reconstructed images, the structural similarity index measure (SSIM) was computed for each image pair. 
The SSIM scores \citep{Zhou2004} displayed below the reconstructed images in \autoref{fig5} are designed to evaluate similarity in a manner consistent with human visual perception, where values closer to 1 indicate greater similarity between two images.

In approach (C), the same ConvAE was first pretrained for parameter initialisation in the DEC method. The same network architecture and training parameters as in approach (B) were adopted for this pretraining stage. 
Following pretraining, cluster centroids were initialised in the latent space using \textit{k}-means clustering, with the number of clusters specified as 7. The convergence tolerance, set at $0.001$, was defined as the threshold for stopping training during the parameter optimisation.
Training was terminated when the proportion of samples exhibiting a change in cluster membership fell below this threshold. 
The parameter $\alpha$ denotes the degree of freedom in Student’s t-distribution used to calculate the soft assignment distribution of samples. It controls the dispersion of the soft labels. Theoretically, $\alpha$ can take any positive value. Larger values (typically $\alpha > 10$) cause the distribution to converge towards a Gaussian shape, while smaller values lead to sharper cluster boundaries.
At $\alpha = 1.0$, the Student’s t-distribution becomes equivalent to a Cauchy distribution, providing heavier tails than a Gaussian distribution.
In this study, we set $\alpha = 1.0$ following the original DEC framework established by \citet{Xie2016}. 
During the clustering optimisation process, the target distribution was updated every 10 epochs to ensure the stability of the cluster assignments. The DEC training converged at epoch 100, at which point the change in cluster membership reached the predefined tolerance threshold.

\begin{figure*}
    \centering
    \includegraphics[width=1\linewidth]{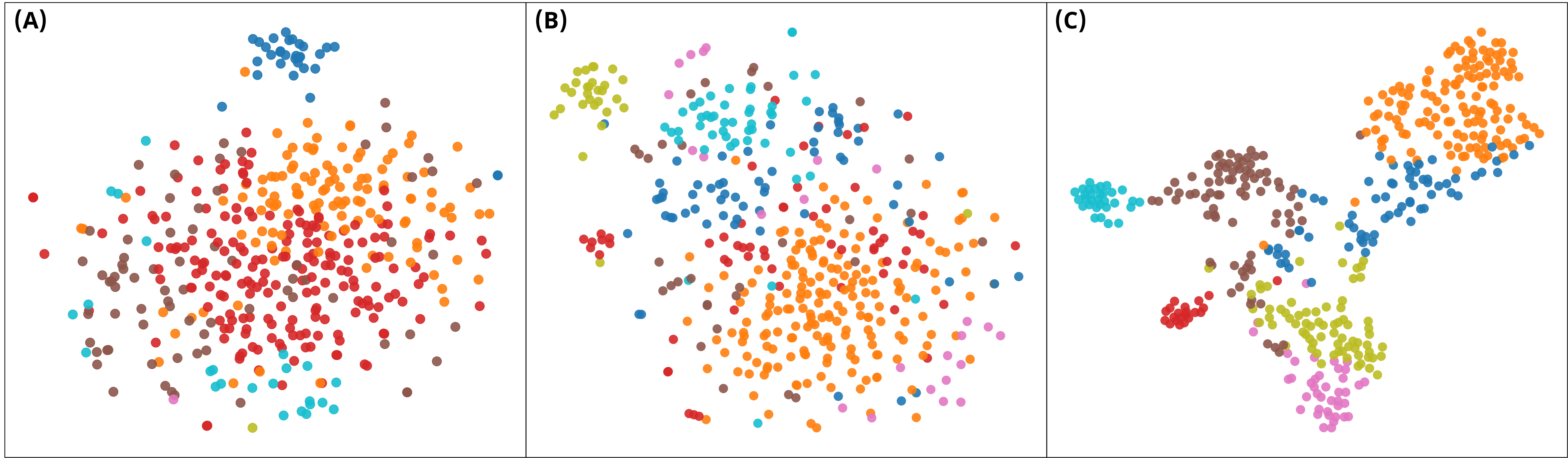}
    \caption{t-SNE visualisation of unsupervised clustering results obtained through three different approaches: (A) PCA + \textit{k}-means; (B) ConvAE + PCA + \textit{k}-means; (C) DEC. Cluster colours in each panel are assigned arbitrarily; therefore, the same colour across different panels does not signify the same glitch morphology.}
    \label{fig6}
\end{figure*}

\subsection{Performance Evaluation and Discussion}\label{sec:result2}

\autoref{fig6} illustrates the clustering results obtained by using approaches (A), (B), and (C) on the test set. 
Since t-SNE is designed to preserve local neighborhood relationships in high-dimensional data, clusters that appear as compact and well-separated groups in the low-dimensional projection indicate that the embedded feature representations effectively capture group structures in the data. 
In contrast, more diffuse or overlapping distributions suggest that the embedded vectors fail to capture the discriminative features of the input data, resulting in poor clustering performance. 
The compressed feature vectors were grouped into seven clusters, with each sample coloured according to its predicted cluster assignment. 
Due to the unsupervised nature of the task, however, the cluster labels or colours have no intrinsic semantic meaning. 
Therefore, it should be noted that the ordering of the cluster labels is arbitrary, and identical colours in panels (A), (B), and (C) in \autoref{fig6} do not correspond to the same clusters; they only indicate cluster membership within each visualisation. 

A comparative analysis of the three visualisations reveals that approach (A) manifests clusters that are relatively sparse and intermingled, indicating limited separation in the PCA-based feature space.  The incorporation of ConvAE features in approach (B) results in more compact groupings within the same clusters, suggesting that non-linear feature extraction enhances the latent representation. Approach (C) yields the most distinctly separated and compact clusters. To provide a quantitative comparison of the cluster separation and cohesion among different methods, we computed the silhouette score, with the results summarised in \autoref{table4}. The silhouette score, defined as the average of the silhouette values across all samples, is an internal evaluation metric for the predicted labels generated by each clustering method. The results indicate that approach (A), PCA + \textit{k}-means, yielded the lowest clustering quality, followed by approach (B), ConvAE + PCA + \textit{k}-means. In contrast, approach (C), DEC, obtained the notable highest clustering quality, indicating that its clusters are more compact and distinctly separated, which is consistent with the visualisation of (C) in \autoref{fig6}. This superiority can be attributed to the fact that, unlike other approaches that directly apply \textit{k}-means clustering to reduced vectors, the DEC framework jointly optimises the latent space representation and the clustering assignments, thereby refining the cluster structure during training. This quantitative advantage is consistently reflected in the two-dimensional $t$-SNE visualisations, where the glitch categories are isolated into well-defined, easily identifiable clusters.

\begin{table}
\centering
\begin{tabular}{cc}
\hline
Method                    & Silhouette Score \\ \hline
PCA + \textit{k}-means             & 0.0097            \\
Conv   AE + PCA + \textit{k}-means & 0.0503            \\
DEC                       & \textbf{0.3042}            \\ \hline
\end{tabular}
\caption{Clustering quality comparison among the three unsupervised approaches, with the best results highlighted in bold.}
\label{table4}
\end{table}

The clustering performance of the three methods compared with respect to ground-truth labels is summarised in \autoref{table5}. The baseline method (A), which combines PCA with \textit{k}-means clustering, achieved the lowest NMI and ARI scores among the three approaches. 
This result reflects the limitation of linear PCA embeddings in sufficiently capturing the intrinsic characteristics of spectrogram images.  
This limitation can be explained with the cumulative explained variance ratio (CEVR) which represents the proportion of total variance captured by the first $n$ principal components during PCA embeddings. Large values of CEVR indicate that the essential morphological features of the input images are preserved well within the lower-dimensional subspace. 
In this study, CEVR of the 50-dimensional PCA vectors was found to be only 31.29\%, supporting the observed degradation in clustering performance.  
In contrast, approaches that incorporate deep learning-based feature extraction demonstrate substantial performance improvements. 
Approach (B), which combines ConvAE with the baseline method, achieved the highest agreement with the ground-truth label distributions, as indicated by the ARI score. 
This suggests that the non-linear embeddings learned by the ConvAE provide more discriminative latent representations for clustering. 
Approach (C), which adopts the DEC algorithm, also outperformed the baseline method by a large margin. 
In particular, approach (C) achieved superior performance in terms of the NMI metric, meaning DEC conserves the intrinsic structure in glitch distributions well. 
These findings emphasize the limitations of PCA embeddings for glitch data analysis and highlight the importance of deep learning-based approaches for effective feature extraction during dimensionality reduction.

\begin{table}
\centering
\begin{tabular}{ccc}
\hline
Method                    & NMI    & ARI    \\ \hline
PCA + \textit{k}-means             & 0.2181 & 0.1711 \\
Conv   AE + PCA + \textit{k}-means & 0.3395 & \textbf{0.4361} \\
DEC                       & \textbf{0.3673} & 0.2981 \\ \hline
\end{tabular}
\caption{Clustering performance comparison with the ground-truth labels among the three approaches, with the best results for each metric highlighted in bold.}
\label{table5}
\end{table}

\begin{figure*} 
    \centering
    \includegraphics[width=0.8\linewidth]{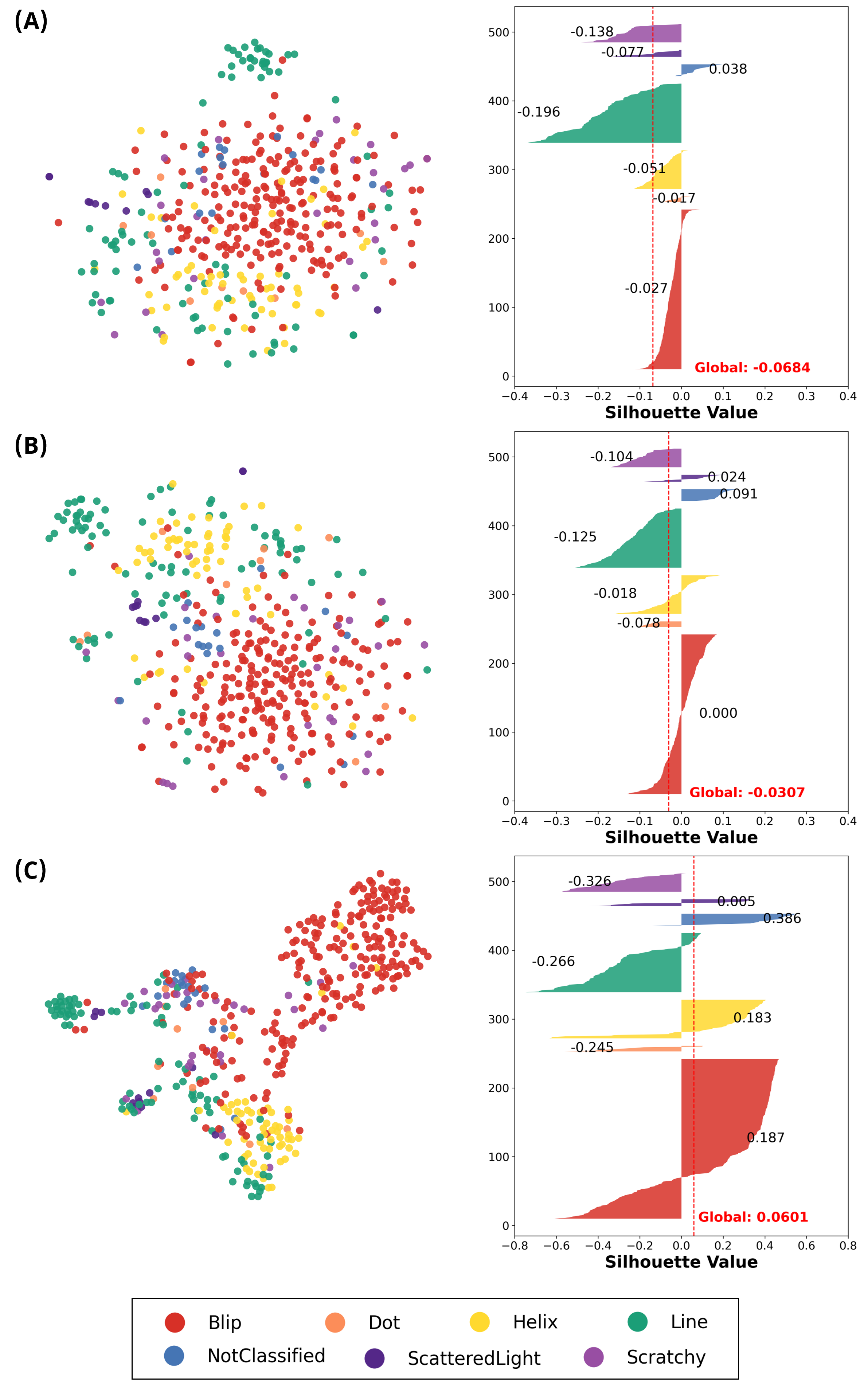}
    \caption{(Left) t-SNE visualisation of samples for each approach, where colours represent the seven glitch classes shown below. (Right) Silhouette values for ground-truth labels; the x- and y-axes denote individual sample silhouette values and data indices, respectively. The red dashed line indicates the global silhouette score, and the black numeric labels indicates the mean silhouette value across each class. (A) PCA + \textit{k}-means; (B) ConvAE + PCA + \textit{k}-means; (C) DEC.}
    \label{fig7}
\end{figure*}

\autoref{fig7} presents a t-SNE visualisation of the human-annotated ground-truth labels of the original data, showing how the glitch classes are distributed in 2D space. 
To examine the separation of the different glitch classes quantitatively in this visualisation, silhouette values were computed for each sample in the embedded latent space and presented in the right panel. 
Note that the silhouette values on \autoref{fig7} serve only as a reference for examining the distribution of the input data in the embedded latent space and do not directly represent the clustering capability of the models.
The average silhouette values of all glitch samples in the three methods (A), (B) and (C) are close to zero, indicating that the boundaries between ground-truth classes are ambiguous or not clearly separated in the latent space.
This reflects the inherent characteristics of the KAGRA O3GK glitch dataset and reveals that the input data itself exhibits a deficiency in intra-cluster cohesion. 
As a result, low silhouette values can still be observed even when clustering results show good agreement with the ground-truth labels.
Consistent with the comparison of silhouette scores between predicted labels (shown on \autoref{table4}), calculating the silhouette score of the ground-truth labels also reveals a slight improvement in method (B) compared to the baseline (A), followed by a more pronounced increase in method (C).
Consequently, DEC produced the most distinctly separated clusters in the latent space, leading to clearer group structures in the low-dimensional representation.
This enhanced inter-class separation facilitates qualitative interpretation of glitch distribution on t-SNE visualisation, highlighting the advantage of DEC as a potential tool for future data clustering. 

 Across all three investigated approaches that mapped our dataset into the low-dimensional latent space, the line and scratchy classes demonstrated a tendency to disperse rather than to form well-defined isolated clusters on the t-SNE projections. 
 This structural scatter indicates substantial intra-class variability, suggesting that samples with identical ground-truth labels can possess notable morphological variations that hinder the model from extracting a highly consistent feature representation.
 
Crucially, regarding the newly introduced "Not Classified" category, characterised by a lack of distinct transient structures, the data points exhibited the highest silhouette scores for all three approaches, implying a relatively strong intra-cluster cohesion in the latent space. 
However, cross-referencing this mapping with the predicted label distributions in \autoref{fig6} reveals that the unsupervised models did not isolate these samples into a dedicated independent cluster. 
Instead, the predicted labels for these features were separated across multiple clusters or combined into another cluster. 
This finding highlights a key area for future architectural refinement in unsupervised glitch clustering.

\section{Conclusion}

This study conducts an unsupervised clustering analysis of glitch spectrograms obtained from gravitational wave detectors by applying three different machine learning methods. 
Specifically, deep learning-based feature extraction is introduced to enhance the clustering of glitch events without relying on extensive manual annotations.
These glitch events identified through Hveto during the O3GK run of the KAGRA detector were transformed into spectrograms, and the pre-processed spectrogram images were used as training data for unsupervised learning.
Three unsupervised approaches were systematically compared: a baseline method combining PCA and \textit{k}-means clustering; a ConvAE + PCA + \textit{k}-means method using deep learning-based feature extraction followed by PCA and \textit{k}-means clustering; and a DEC algorithm that optimises deep learning-based feature embedding and clustering simultaneously. 
The subsequent comparative analysis revealed the limitations of PCA-based linear dimensionality reduction in capturing meaningful features, while deep learning-based approaches provided more representative feature embeddings for clustering.
Notably, the ConvAE + PCA + \textit{k}-means framework achieved the highest structural pairwise agreement with the ground-truth distribution, yielding a peak ARI of $0.4361$. 
The DEC algorithm maximized shared mutual information, achieving the highest NMI score of $0.3673$. 
Although the DEC algorithm achieved a lower ARI compared to the ConvAE + PCA + \textit{k}-means, it yielded more distinct and compact clusters within the latent space, demonstrating its advantage for visual interpretation.

While a subset of annotated data was used in this study to evaluate the model's validity, label information was strictly excluded from the training phase and was only used for quantitative evaluation. 
The findings demonstrate the efficacy of deep unsupervised frameworks in bridging the gap between raw instrumental data and meaningful categorisation.
Future work should further enhance the discriminative power of the unsupervised clustering. 
This could be achieved by incorporating advanced algorithms to refine the latent space embedding, ensuring that even morphologically similar or subtle glitch variations are captured with higher precision.
Subsequent research will extend this study to include clustering frameworks that do not require a predefined number of clusters, enabling the fully automated clustering of unlabelled glitch populations without human intervention. 
These adaptive deep learning methods will facilitate the discovery of unknown glitch morphologies, providing the scalable and annotation-free glitch classification pipelines required for next-generation gravitational wave observatories.

\section*{Appendix}
\label{sec:appendix}

\autoref{table2} summarises the architecture of the ConvAE used to map each input spectrogram to a low-dimensional latent representation. The same encoder--decoder architecture, without modification, was reused to construct the DEC model, with the decoder discarded after pretraining and the encoder fine-tuned jointly with the clustering objective (see \autoref{sec:method1}). This appendix describes the notation used in \autoref{table2} and each process in detail.

The encoder consists of three convolution--max-pooling blocks that progressively reduce the spatial resolution of the input from $224 \times 224$ to $28 \times 28$  while increasing the number of channels from 4 to 128, followed by a fully-connected layer that compresses this feature map into a 64-dimensional latent vector. 
The decoder mirrors this structure in reverse, using a fully-connected layer and three upsampling--convolution blocks to reconstruct a $(4, 224, 224)$ tensor matching the shape of the original input. 
A single input spectrogram is represented as a 3-dimensional tensor of shape $(C, H, W)$, where $C$ is the number of feature channels (four in this work), and $H$ and $W$ denote the spatial height and width ($224 \times 224$), respectively.

In \autoref{table2}, the output entry is expressed as a 4-dimensional tensor of shape $(M, C, H, W)$, to account for the mini-batch size, $M$. During training, rather than updating the network weights after processing every single spectrogram, the optimiser processes a small batch of $M$ spectrograms simultaneously. The optimiser then averages the resulting gradients over the batch before updating the weights. This mini-batch strategy reduces the variance of the gradient estimate relative to using a single sample, while remaining computationally far cheaper than using the entire training set at once. $M$ is a user-defined hyperparameter summarised in \autoref{table3}.
Layers marked with the superscript $B$ are followed by a batch normalisation operation. For each feature channel, batch normalisation rescales the layer's activations, computed across the current mini-batch of $M$ samples, to have approximately zero mean and unit variance. This has two practical effects: it mitigates the tendency of activation statistics to drift as the weights of earlier layers are updated during training, and it empirically allows the use of larger learning rates and faster, more stable convergence during optimisation. Batch normalisation is applied only during the convolutional feature-extraction stages of the network.

\begin{table}
    \centering
        \begin{tabular}{lll}
            \toprule
            \textbf{Encode Layer} & \textbf{Output} & \textbf{Activation} \\
            \midrule
            Input           & (M, 4, 224, 224)           & \\
            Convolution     & (M, 32, 224, 224)$^{\mathrm{B}}$ & ReLU \\
            Max-Pooling     & (M, 32, 112, 112)          & \\
            Convolution     & (M, 64, 112, 112)$^{\mathrm{B}}$  & ReLU \\
            Max-Pooling     & (M, 64, 56, 56)            & \\
            Convolution     & (M, 128, 56, 56)$^{\mathrm{B}}$  & ReLU \\
            Max-Pooling     & (M, 128, 28, 28)           & \\
            Fully-Connected & (M, 64)                    & \\
            \midrule
            \textbf{Decode Layer} & \textbf{Output} & \textbf{Activation} \\
            \midrule
            Input           & (M, 64)                    & \\
            Fully-Connected & (M, $128\times28\times28$) & \\
            Upsampling      & (M, 128, 56, 56)           & \\
            Convolution     & (M, 64, 56, 56)$^{\mathrm{B}}$             & ReLU \\
            Upsampling      & (M, 64, 112, 112)          & \\
            Convolution     & (M, 32, 112, 112)$^{\mathrm{B}}$ & ReLU \\
            Upsampling      & (M, 32, 224, 224)          & \\
            Convolution     & (M, 4, 224, 224)           & Sigmoid \\
            \bottomrule
        \end{tabular}  
    
     \caption{Proposed architecture of the ConvAE model used for mapping input spectrograms to low-dimensional latent vectors. $B$ and $M$ denote batch normalisation and the mini-batch size, respectively.}  
     \label{table2}
\end{table}

\section*{Acknowledgements}

Kwak K. was also supported by Institute of Information \& communications Technology Planning \& Evaluation (IITP) grant funded by the Korea government (MSIT) (No.RS-2020-II201336, Artificial Intelligence Graduate School Program (UNIST))

\section*{Data Availability}
 
The data used in this study are publicly available from the Gravitaional Wave Open Science Center at \url{https://gwosc.org/O3/O3GK/}. Strain data can be accessed and downloaded from this site. Auxiliary channel data used in the Hveto analysis are not publicly available.



\bibliographystyle{mnras}
\bibliography{reference} 








\bsp	
\label{lastpage}
\end{document}